# Giant moiré trapping of excitons in twisted hBN

Yanshuang Li[1,2], Xiuhua Xie[1,*], Huan Zeng[1,2], Binghui Li[1], Zhenzhong Zhang[3], Shuangpeng Wang[4], Jishan Liu[5], Dezhen Shen[1,*]

[1]State Key Laboratory of Luminescence and Applications, Changchun Institute of Optics, Fine Mechanics and Physics, Chinese Academy of Sciences, No. 3888 Dongnanhu Road, Changchun, 130033, People's Republic of China

[2]University of Chinese Academy of Sciences, Beijing 100049, People's Republic of China

[3] School of Microelectronics, Dalian University of Technology, Dalian, 116024, P. R. China

[4] MOE Joint Key Laboratory, Institute of Applied Physics and Materials Engineering and Department of Physics and Chemistry, Faculty of Science and Technology, University of Macau, Macao SAR 999078, P. R. China

[5] Center for Excellence in Superconducting Electronics, State Key Laboratory of Functional Materials for Informatics, Shanghai Institute of Microsystem and Information Technology, Chinese Academy of Sciences, Shanghai 200050, China

*Corresponding authors. Email: xiexh@ciomp.ac.cn; shendz@ciomp.ac.cn

# Abstract

Excitons in van der Waals (vdW) stacking interfaces can be trapped in ordered moiré potential arrays giving rise to attractive phenomenons of quantum optics and bosonic many-body effects. Compare to the prevalent transition metal dichalcogenides (TMDs) systems, due to the wide bandgap and low dielectric constant, excitons in twist-stacked hexagonal boron nitride (hBN) are anticipated trapped in deeper moiré potential, which enhances the strength of interactions. However, constrained by the common low detectivity of weak light-emitting in the deep-ultraviolet (DUV) bands, the moire excitons in twist-hBN remain elusive. Here, we report that a remarkable DUV emitting band (peak located at ~260 nm) only emerges at the twisted stacking area of hBN, which is performed by a high collection efficiency and spatially-resolved cathodoluminescence (CL) at room temperature. Significant peak redshifting contrast to defect-bound excitons of bulk hBN indicates the giant trapping effects of moiré potential for excitons. The observation of deeply trapped excitons motivates further studies of bosonic strongly correlation physics based on the twist-hBN system.

**Introduction**

Moiré superlattices with a nanoscale periodic trapping potential arising from lattice mismatch or twist-angle between stacked monolayers have triggered a boom in research of condensed matter quantum simulators[1-3]. Besides the proven charge order quantum transition in moiré bilayers, such as generalized Wigner crystals[4], Mott insulators[5], to name a few, excitons (electron-hole bound by Coulomb interactions) coherence phases and nonlinearity are upcoming based on bilayers transition metal dichalcogenides (TMDs)[6-8]. The deep moiré trapping effect of excitons has become a significant basis for many-boson quantum phenomena in the future including Boson-Hubbard models, Bose-Einstein condensate, and so on.

Hexagonal boron nitride (hBN), acting as an ideal encapsulator or spacer for two-dimensional (2D) materials heterostructures, also has abundant optoelectronic properties from deep-ultraviolet (DUV) to mid-infrared range[9]. Ultra-wide energy bandgap (> 6 eV) admits a store of defect levels, which induces purely single-photon sources at room-temperature[10]. As natural hyperbolic materials, it hosts phonon-polaritons supporting near-field optical applications[11]. Notably, compared with TMDs, excitons are compact (Bohr radius around 8 Å, less than nanoscale moiré pattern) in hBN with larger binding energy (several hundred meV)[12-14]. This makes excitons can be trapped at high symmetry of moiré superlattices, within a larger range of twist-angles, which determine the moiré lattice constants. Stacking and twisting of layered hBN have significantly affected the shifting of band edge[15], suggesting larger potential energy offset between local atomic registries, also means deeper moiré potential. Hitherto, subjected to the technical challenges in DUV bands, especially in the weak emitting collection and spatial resolution, optical signatures of moiré excitons in twisted-hBN have remained unexplored.

Herein, we report an emerging DUV emitting peak located around 260 nm in twisted-hBN structures via advanced cathodoluminescence (CL) measurement. High spatial-resolved hyperspectral imaging demonstrates that this new peak only existed in the stacking area. This kind of large energy difference to defect-bound excitons is improbable from phonon-replicas. Moreover, charging and strain effect inducing peak redshifting have been excluded by measuring the carbon-coated and wrinkle sample. These findings support giant potential trapping of excitons in twisted-hBN, which is up to 807 meV based on emission spectrum. Our work motivates further studies of moiré physics in hBN systems, for instance, bosonic many-body, flat-bands, moiré ferroelectricity, etc.

**Experiment details**

The few-layers hBN mechanical exfoliated on the polydimethylsiloxane (PDMS) stamps by standard Scotch-tape method from bulk hBN (HQ graphene, Netherlands). Targeted layered areas and tearing edges were identified by optical microscopy. Twisted-hBN was obtained by stacking each layer on sapphire substrate in turn via heating PDMS at 85 °C 20 mins. A clean interface was achieved in our dry transfer approach by avoiding excessive use of polymers as well as solution cleaning processes. The twist-angle between hBN sheets was determined by both edges-aligned and oriented-wrinkles (along armchair direction)[16], which is 67.96° visually, while, taking into account $D_{6h}$ point group, we label twist-angle in our work is 7.96°.

For achieving the higher collection efficiency and spatial resolution in DUV range from moiré excitons emission, an upgraded CL measurement (Attolight AG., Switzerland) was carried out. The CL system makes a high numerical aperture (NA=0.72) achromatic reflective integrated within the objective lens of a scanning electron microscope (SEM). The focal plane of the optical lens is matched with the optimal working distance of SEM, so that matches the field of view of both optical and electron microscopy. It works in the energy of low electron beam (3 keV) meets higher spatial resolution CL hyperspectral images. Emissions was coupled to the free space, wavelength resolved by a monochromator (Czerny-Turner configuration), measured with a UV/Vis detector.

**Results and discussion**

Figure 1a and b show schematic diagrams of moiré exciton and CL objective. There has a large energy difference between interlayer atomic high-symmetry points within the moiré unit cell, which induces excitons to be trapped and localized at moiré potential wells. The tightly integrated optical reflective lens makes sufficient collection efficiency even under relatively weak excitation intensity, which is particularly useful for the detection of excitons luminescence in the DUV range. As shown in Fig. 1c, besides defects and impurities-related bands in near-UV and visible wavelength, twisted hBN exhibits a distinct DUV emission band ($\lambda < 280$ nm) with multi-peaks at room temperature. Accurately fitted spectrum lines (peak wavelength) of twisted area hBN have been listed in Table 1. The spectrum line 221.803 (S1) and 228.425 (S2) nm are BN near banggap edge excitonic emission peaks, which are produced by excitons bound with defects[17-19]. The spectral line 297.877 (S4), 311.967 (S5), 325.310 (S6), 339.446 (S7), 542.413 (S8), 555.124 (S9) and 581.000 nm (S10) are impurities and defects related emission peaks[17, 20]. Among the origin of these peaks, S4 to S7 may be derived carbon and oxygen incorporation, still controversial[17]. And the S8 to S10 emission peaks are derived from different boron or nitrogen vacancy related defects[21,22]. Notably, there is a remarkable emission peak, 260.302 nm (S3), which is seldom presented in previous reports. Here we argue that it comes from the moiré excitonic emission peak (see below for further discussion). With the effect of moiré potential, near bandgap edge excitons are trapped in the moiré potential well, and then excitonic emission peaks occur a large redshift. Compared with the photon energy of S1, the energy redshift of moiré exciton emission peak S3 is about 807 meV for twisted angle hBN at room temperature. Such moiré potential in twisted hBN is deeper than twisted TMDs systems[23, 24], which will be conducive to control of the excitons correlation and achieving artificial excitonic many-body phases.

Firstly, to confirm the distribution area of S3, a highly spatial-resolved electron beam scanning mode of CL has been utilized. The twisted area of stacked hBN has been surrounded by a red dotted line in the image of SEM, as shown in Fig. 2(a). Owing to the penetration of the electron beam, the overall stronger emitting intensity is concentrated in the single area, which mainly comes from the Farbe-center of the sapphire substrate, also serving as calibration of peaks, as can be seen from the panchromatic image, Fig. 2(b). Colorized bands images can be found in Supplementary Fig. S1. Five localized spectra, positions labeled in Fig. 2(b), have been taken in Fig. 2(c). Noticeably, the S3 is only present at the twisted area while not at the single area. This suggests that S3 is caused by the twisting effect of layered hBN. Additionally, S3 has a larger redshift from the band-edge, which is unattainable energy trapping deepness by the ordinary defects bound excitons in hBN[19, 25]. This indicates that the S3 is probably come from the moiré trapped excitons in superlattices.

Charging effects induced by the electron beam of CL may impact emission peak position. Here, we have reduced it further by surface treatment of carbon-coating after initial measurement. The carbon has conductivity, avoiding the surface charges accumulation, while with no emission signals in the wavelength range of DUV. As shown in Fig. 3(c), localized spectra extracted from Fig. 3(b), marked 1 to 3, corresponding to the twisted area, single area, and sapphire, respectively, there are no observable peak-shiftings vs Fig. 2(c), which demonstrated that the adopted low electrons excitation intensity (3 keV) does not produce a significant charge effect. Consistently, the S3 emission band is still only emerged in the twist area, definitely. The results show that the arising of S3 does not derive from the surface charging effect.

Charging effects induced by the electron beam of CL may impact emission peak position. Here, we have reduced it further by surface treatment of carbon-coating after initial measurement. The carbon has conductivity, avoiding the surface charges accumulation, while with no emission signals in the wavelength range of DUV. As shown in Fig. 3(c), localized spectra extracted from Fig. 3(b), marked 1 to 3, corresponding to the twisted area, single area, and sapphire, respectively, there are no observable peak-shiftings vs Fig. 2(c), which demonstrated that the adopted low electrons excitation intensity (3 keV) does not produce a significant charge effect. Consistently, the S3 emission band is still only emerged in the twist area, definitely. The results show that the arising of S3 does not derive from the surface charging effect.

On account of the flexibility of PDMS, there have several unintentionally produced wrinkles of top-layer hBN during the stacking process, as shown in SEM images of Fig. 2(a) and Fig. 3(a). The wrinkles (layer folding and protuberant area, Fig. 4(a)) have a conspicuous feature of strain, which also play a role in excitons localization[26]. As can be seen in Fig. 4(c) and Supplementary Fig. S3(b), S1 and S2 show enhanced luminescence intensities at the wrinkles regions on account of the funneling effect on excitons[27]. The DUV emission bands show redshift as strain increases, from flat to fold intersection area, marked in the panchromatic image (1 to 3 in Fig. 4(b)). The peak shifting is about 15 nm between locations 1 and 3, corresponding to ~359 meV, which proves that S3 is not induced by strain effects. Moreover, there are no clear S3 related bands in fold areas, owing to interlayer decoupling results of distance enlarge between stacked layers at this area, which are further support that S3 comes from the interlayer coupling of twisted stacking.

Finally, the CL peaks influence by electron beam current density have been studied via adjusting the field of view (FOV). Under the same beam energy (3 keV), a smaller FOV corresponds to a larger beam current density. As shown in Fig. 5(a), the calibration peak, Farbe-center emission of the sapphire substrate, has no changes with varying FOV from 5 μm to 30 μm, which indicates that under the conditions that we utilized, CL peaks are not affected by changes of electron beam current density. In addition, a thick area of hBN has also been tested. As plotted in Fig. 5(b), there is also no S3 band that existed in the thick area. Therefore, the S3 band cannot be caused by the changing of pump intensity and thickness[18].

In conclusion, by utilizing the high collection efficiency and spatially-resolved CL, we found an emerging S3 emission band only located twisted area hBN at room temperature. Through comprehensive studies of charging effect, strain, pump intensity, and layer thickness, the energy redshift between S1 and S3 is nontrivial. All experimental evidence supports that the S3 related band is coming from the giant trapping of moiré potential (about 807 meV) of excitons in twisted layered hBN. Our findings motivate the future studies of excitonic moiré physics based on twisted hBN at higher temperatures, such as moiré array quantum emission, bosonic many-body models, chirality, and so on.

**Acknowledgments**

The authors thank Attolight for their assistance with CL measurements and fruitful discussions. This research is mainly supported by the National Natural Science Foundation of China (Grant Nos. 11727902 and 62074146). S.P. and X.H. gratefully acknowledge support from Multi-Year Research Grants (MYRG2020-00207-IAPME) from Research & Development Office at University of Macau.

## Competing interests

The authors declare no competing interests.

## Reference:

1. Kennes DM, et al. Moiré heterostructures as a condensed-matter quantum simulator. *Nat Phys* **17**, 155-163 (2021).

2. Wilson NP, Yao W, Shan J, Xu XD. Excitons and emergent quantum phenomena in stacked 2D semiconductors. *Nature* **599**, 383-392 (2021).

3. Tong QJ, Yu HY, Zhu QZ, Wang Y, Xu XD, Yao A. Topological mosaics in moiré superlattices of van der Waals heterobilayers. *Nat Phys* **13**, 356-362 (2017).

4. Regan EC, et al. Mott and generalized Wigner crystal states in WSe2/WS2 moiré superlattices. *Nature* **579**, 359-363 (2020).

5. Li TX, et al. Continuous Mott transition in semiconductor moire superlattices. *Nature* **597**, 350-354 (2021).

6. Ma LG, et al. Strongly correlated excitonic insulator in atomic double layers. *Nature 598*, 585-589 (2021).

7. Gu J, et al. Dipolar excitonic insulator in a moire lattice. *arXiv preprint* **arXiv:210806588** (2021).

8. Zhang L, et al. Van der Waals heterostructure polaritons with moiré-induced nonlinearity. *Nature* **591**, 61-65 (2021).

9. Caldwell JD, Aharonovich I, Cassabois G, Edgar JH, Gil B, Basov DN. Photonics with hexagonal boron nitride. *Nat Rev Mater* **4**, 552-567 (2019).

10. Sajid A, Ford MJ, Reimers JR. Single-photon emitters in hexagonal boron nitride: a review of progress. *Reports on Progress in Physics* **83**, 044501 (2020).

11. Li PN, et al. Hyperbolic phonon-polaritons in boron nitride for near-field optical imaging and focusing. *Nat Commun* **6**, 7507 (2015).

12. Cao XK, Clubine B, Edgar JH, Lin JY, Jiang HX. Two-dimensional excitons in three-dimensional hexagonal boron nitride. **Applied Physics Letters** *103*, 191106 (2013).

13. Cassabois G, Valvin P, Gil B. Hexagonal boron nitride is an indirect bandgap semiconductor. *Nature Photonics* **10**, 262-266 (2016).

14. Rousseau A, et al. Monolayer Boron Nitride: Hyperspectral Imaging in the Deep Ultraviolet. *Nano Lett*, (2021).

15. Zhao XJ, Yang Y, Zhang DB, Wei SH. Formation of Bloch Flat Bands in Polar Twisted Bilayers without Magic Angles. *Phys Rev Lett* **124**, 6 (2020).

16. Chen L, et al. Direct observation of layer-stacking and oriented wrinkles in multilayer hexagonal boron nitride. *2D Mater* 8, 024001 (2021).

17. Watanabe K, Taniguchi T. Far-UV photoluminescence microscope for impurity domain in hexagonal-boron-nitride single crystals by high-pressure, high-temperature synthesis. *npj 2D Mater Appl* **3**, 40 (2019).

18. Schue L, et al. Dimensionality effects on the luminescence properties of hBN. *Nanoscale* **8**, 6986-6993 (2016).

19. Watanabe K, Taniguchi T, Niiyama T, Miya K, Taniguchi M. Far-ultraviolet plane-emission handheld device based on hexagonal boron nitride. *Nature Photonics* **3**, 591-594 (2009)

20. Museur L, Anglos D, Petitet JP, Michel JP, Kanaev AV. Photoluminescence of hexagonal boron nitride: Effect of surface oxidation under UV-laser irradiation. *Journal of Luminescence* **127**, 595-600 (2007).

21. Hayee F, et al. Revealing multiple classes of stable quantum emitters in hexagonal boron nitride with correlated optical and electron microscopy. *Nat Mater* **19**, 534-539 (2020).

22. Mendelson N, et al. Identifying carbon as the source of visible single-photon emission from hexagonal boron nitride. *Nat Mater* **20**, 321-328 (2021).

23. Shabani S, et al. Deep moiré potentials in twisted transition metal dichalcogenide bilayers. *Nat Phys* **17**, 720-725 (2021).

24. Wang X, et al. Moire trions in MoSe2/WSe2 heterobilayers. *Nat Nanotechnol* **16**, 1208-1213 (2021).

25. Schue L, et al. Bright Luminescence from Indirect and Strongly Bound Excitons in h-BN. *Phys Rev Lett* **122**, 067401 (2019).

26. Darlington TP, et al. Imaging strain-localized excitons in nanoscale bubbles of monolayer WSe2 at room temperature. *Nat Nanotechnol* **15**, 854-860 (2020).

27. Koo Y, et al. Tip-Induced Nano-Engineering of Strain, Bandgap, and Exciton Funneling in 2D Semiconductors. *Adv Mater* **33**, 2008234 (2021).

**Figure**

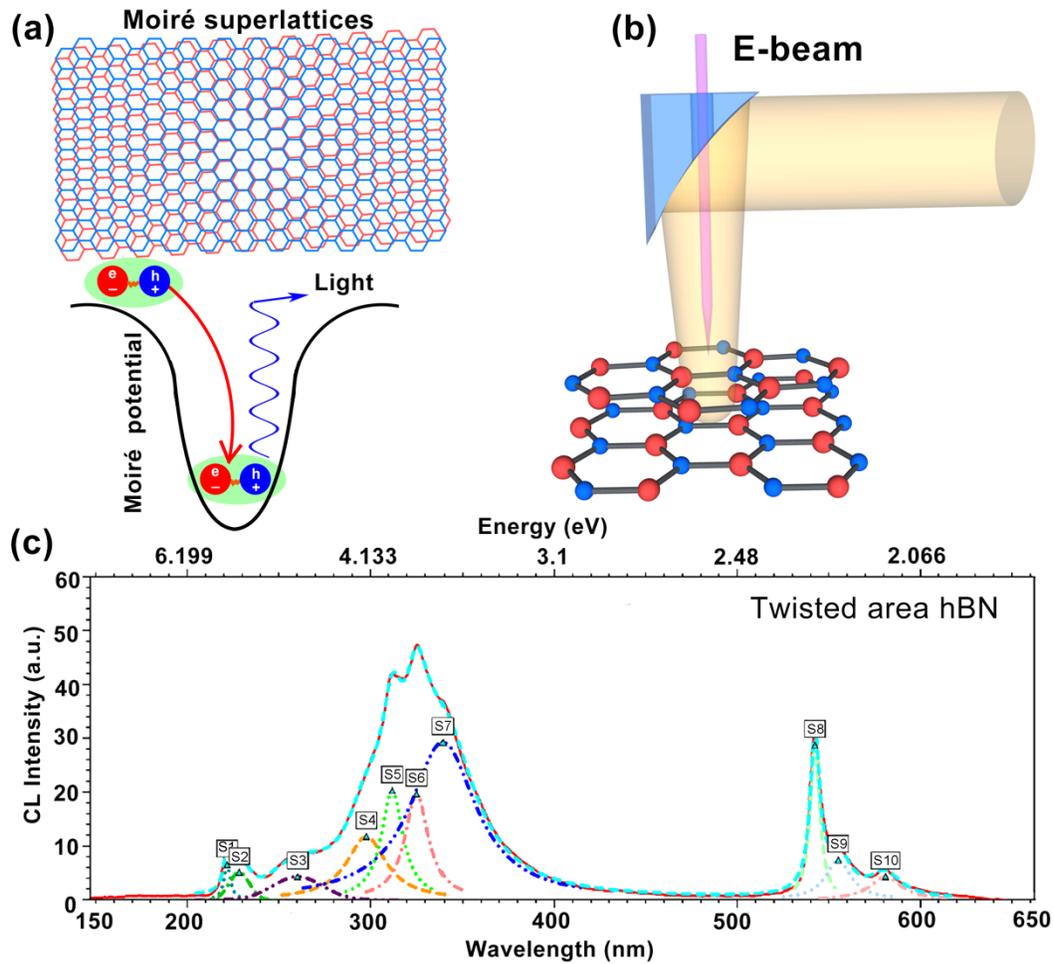

Fig. 1. Cathodoluminescence (CL) spectrum of moiré excitons in twisted hBN. (a) Excitons in moiré trapping. (b) Schematic illustration of cathodoluminescence spectrum. (c) CL and fitted spectrum of twisted area hBN at room temperature.

Table 1. Fitting peaks of twisted hBN

|      | CL Intensity | Wavelength | FWHM |
|------|--------------|------------|--------|
| unit | a. u. | nm | nm |
| S1 | 6.849 | 221.803 | 4.115 |
| S2 | 5.070 | 228.425 | 12.933 |
| S3 | 4.285 | 260.302 | 31.805 |
| S4 | 11.743 | 297.877 | 26.244 |
| S5 | 20.258 | 311.967 | 14.398 |
| S6 | 19.642 | 325.310 | 15.641 |
| S7 | 29.212 | 339.446 | 43.325 |
| S8 | 28.711 | 542.413 | 6.353 |
| S9 | 7.347 | 555.124 | 16.807 |
| S10 | 4.281 | 581.000 | 20.501 |

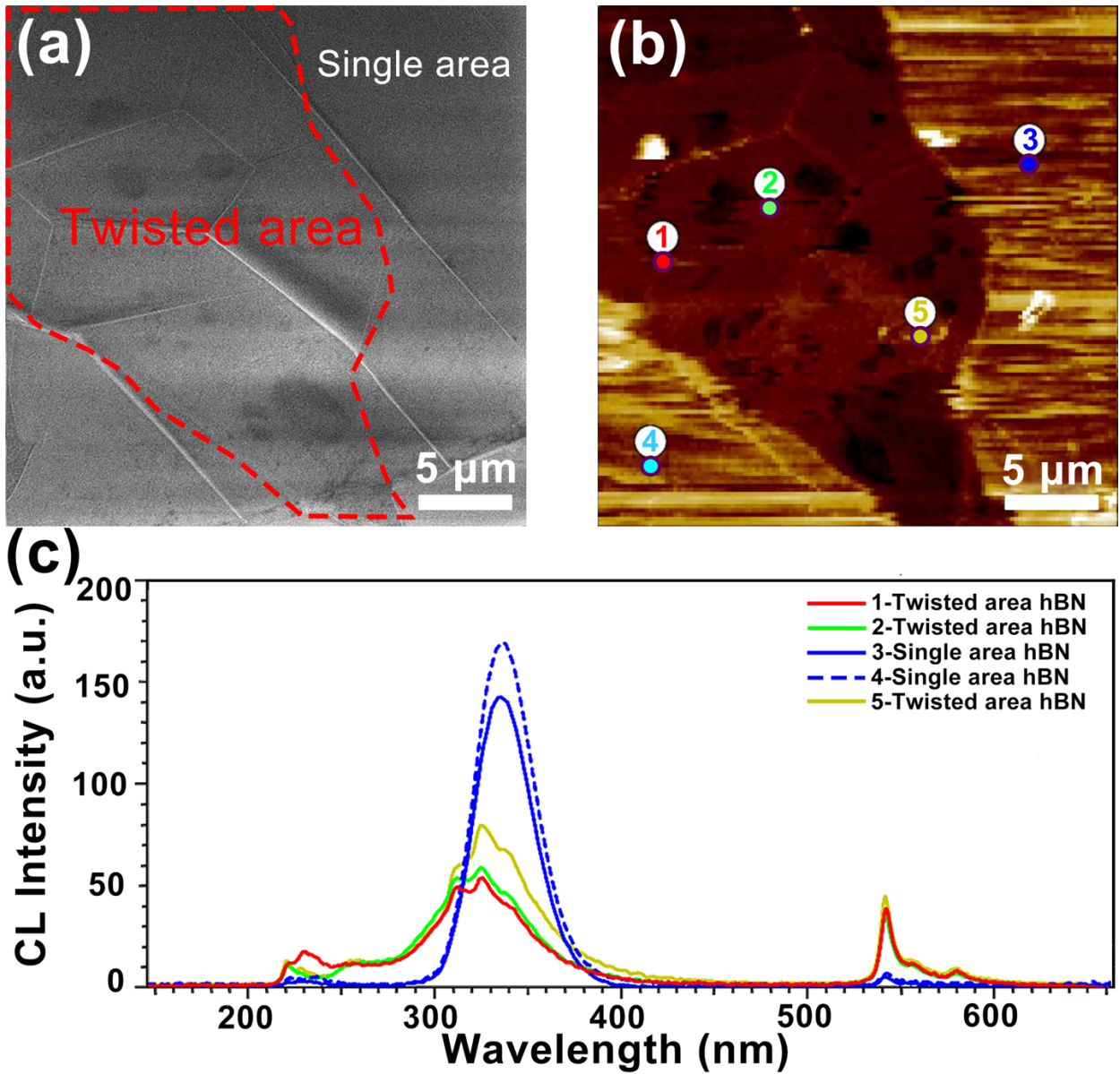

Fig. 2. Scanning electron (SE) images and localized spectrum of hBN. (a) SE images of hBN single and twisted area. (b) Panchromatic SE images. (c) Five positions of hBN CL spectrum. 1, 2 and 5 positions are hBN twisted area. And 3 and 4 positions are BN single layer area.

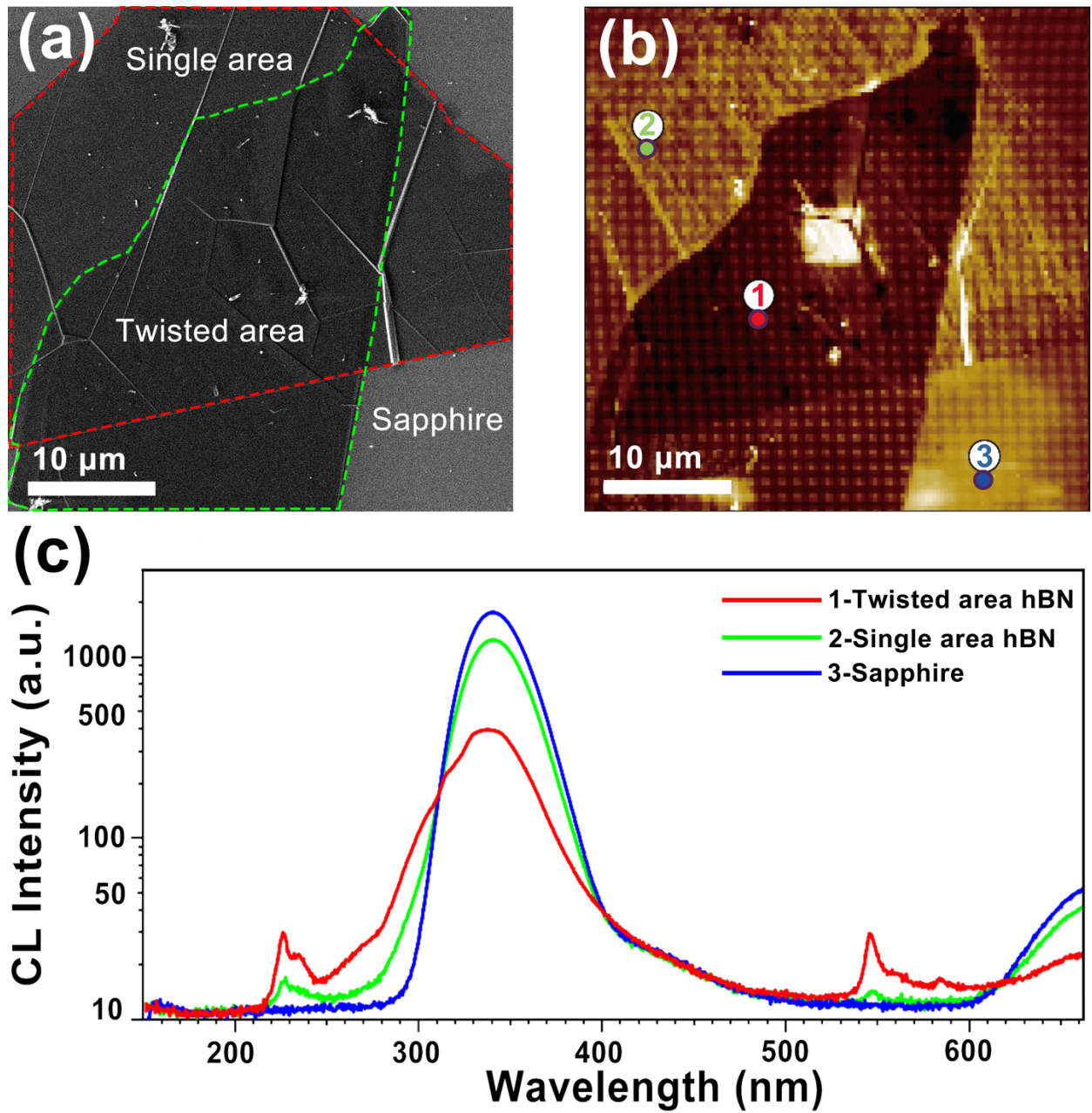

Fig. 3. SE images and localized spectrum of hBN after carbon-coating. (a) SE images of hBN single and twisted area. (b) Panchromatic SE images. (c) Three positions of hBN CL spectrum. 1 position is hBN twisted area. 2 position is BN single layer area. And 3 position is sapphire.

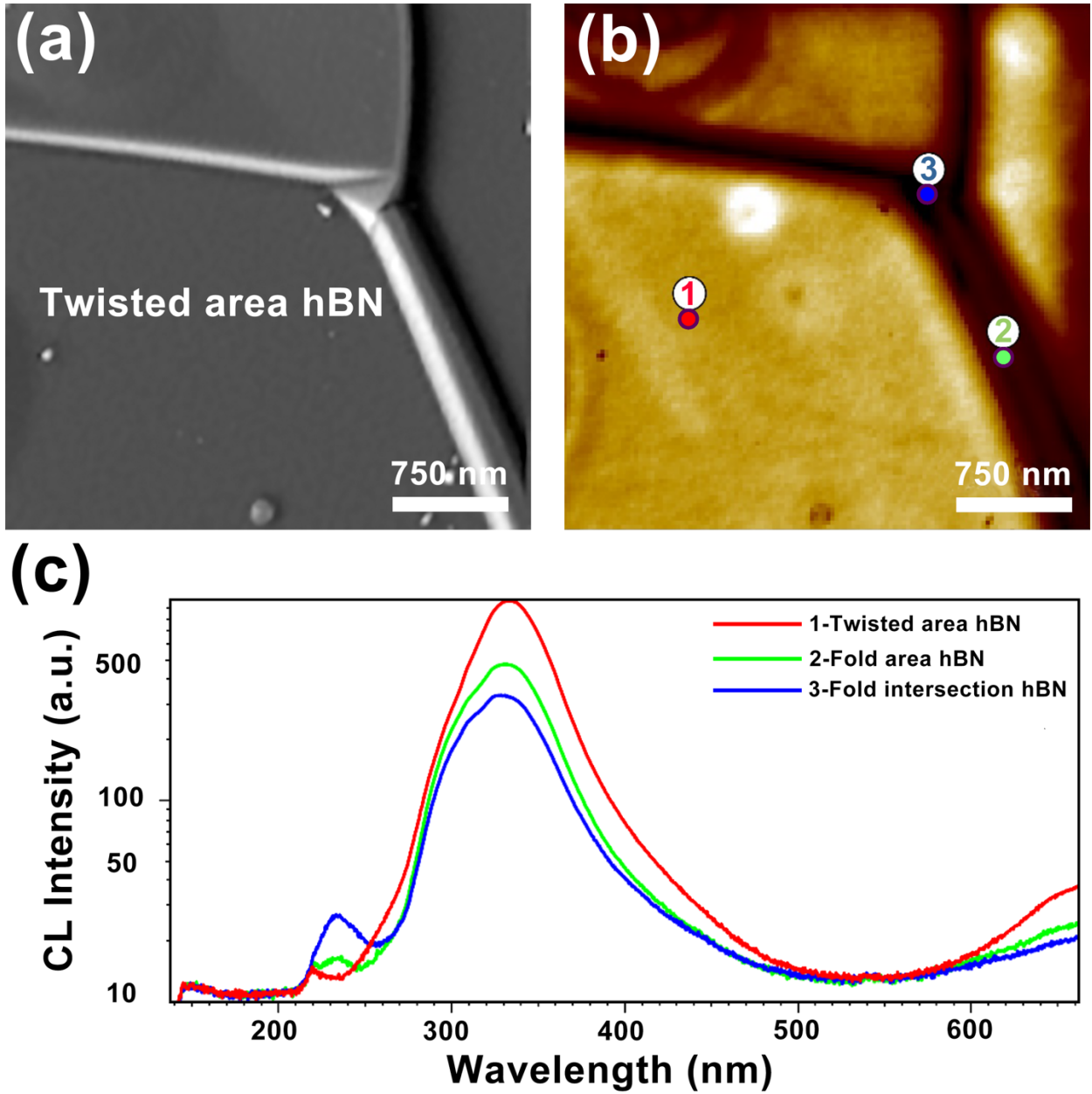

Fig. 4. SE images and localized spectrum of fold area hBN after carbon-coating. (a) SE images of fold twisted area hBN. (b) Panchromatic SE images. (c) Three positions of hBN CL spectrum. 1 position is hBN twisted area. 2 position is hBN fold area. And 3 position is fold intersection area.

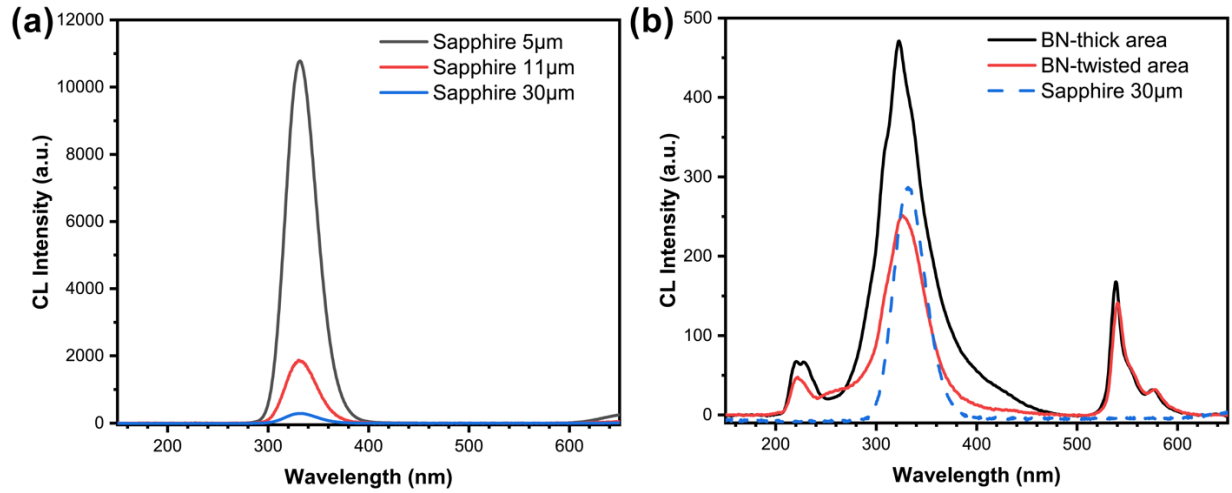

Fig. 5. CL spectrum of sapphire and hBN. (a) Sapphire emission at room temperature various with field of view (electron beam current density). The sapphire emission peak is 332.020 nm. (b) Emission spectrum of hBN thick area, twisted area and sapphire at room temperature.

**Supplementary materials**

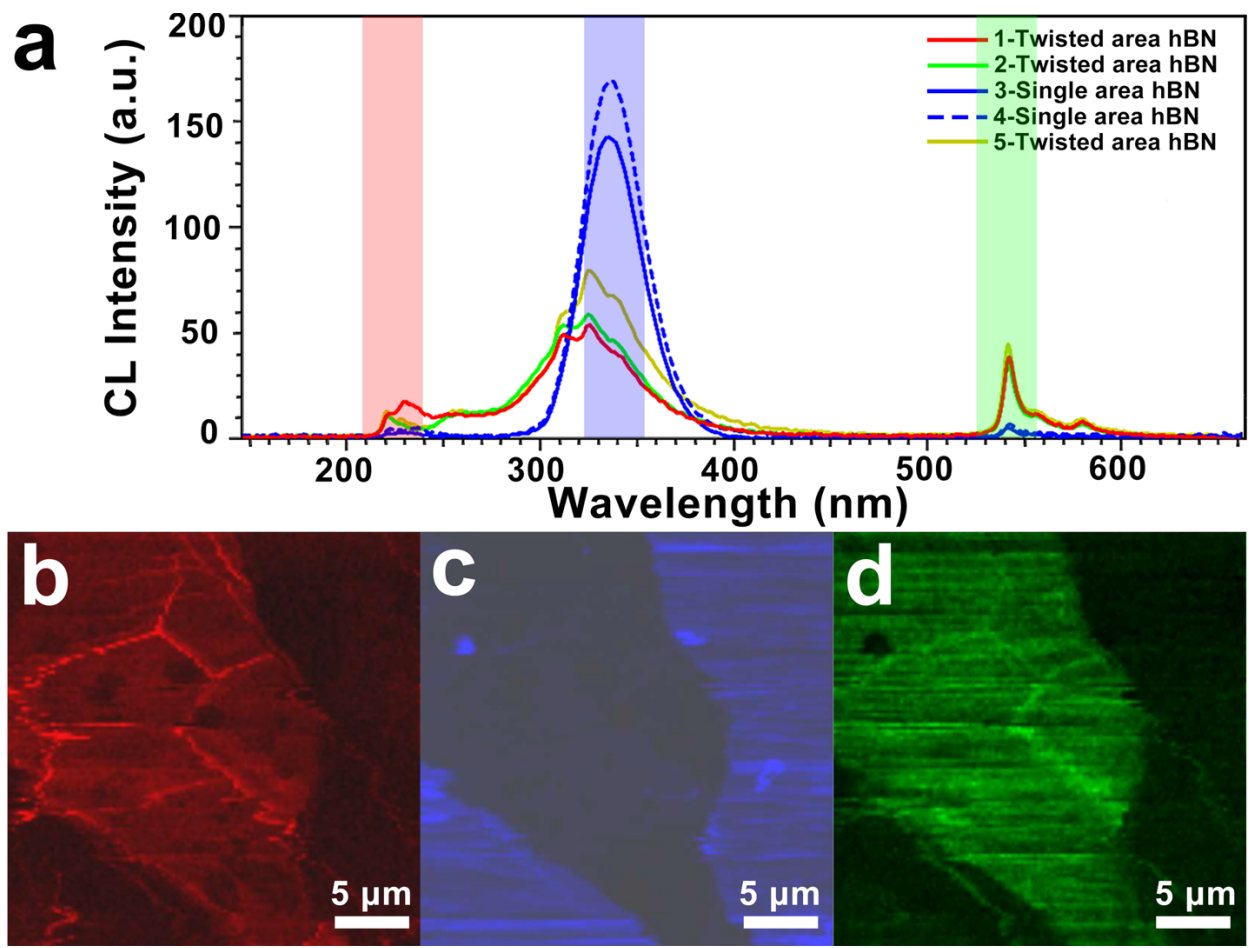

Fig. S1. Colorized bands images of hBN. (a) CL spectra of hBN. (b) Spectral range is red labeled area in Fig S1(a). (c) Spectral range is blue labeled area in Fig S1(a). (d) Spectral range is green labeled area in Fig S1(a).

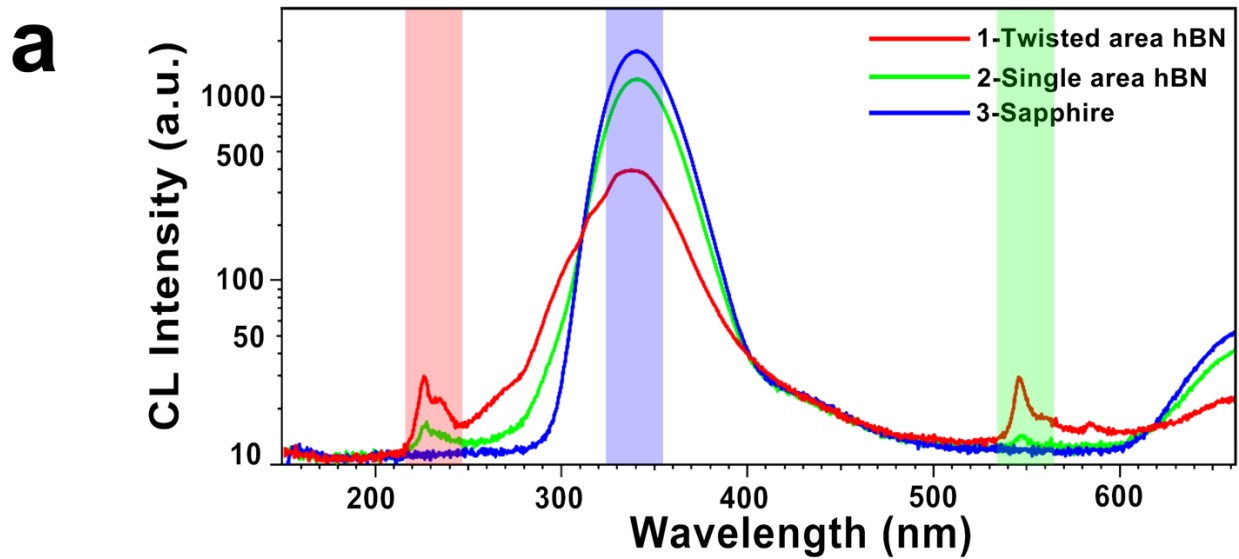
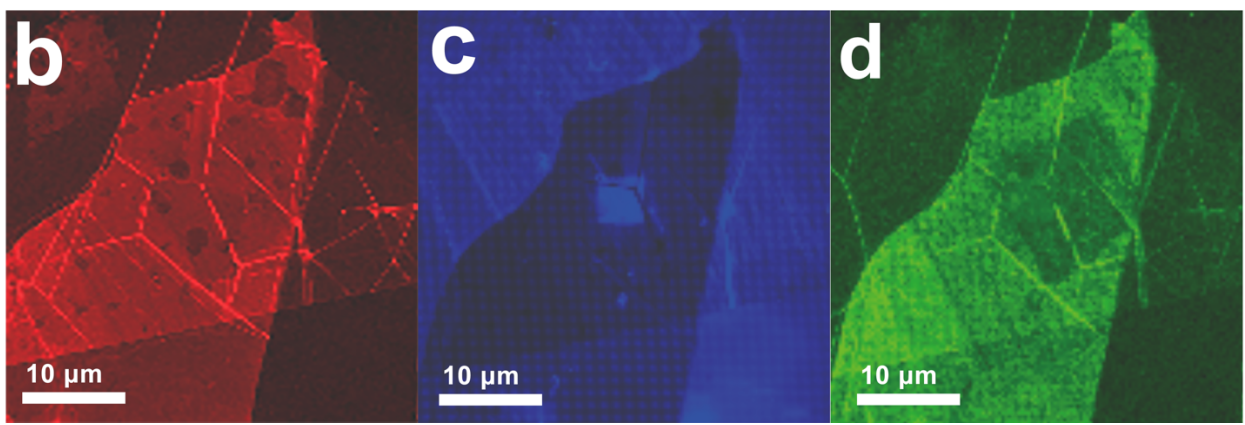

Fig. S2. Colorized bands images of hBN after carbon-coating. (a) CL spectra of hBN. (b) Spectral range is red labeled area in Fig S2(a). (c) Spectral range is blue labeled area in Fig S2(a). (d) Spectral range is green labeled area in Fig S2(a).

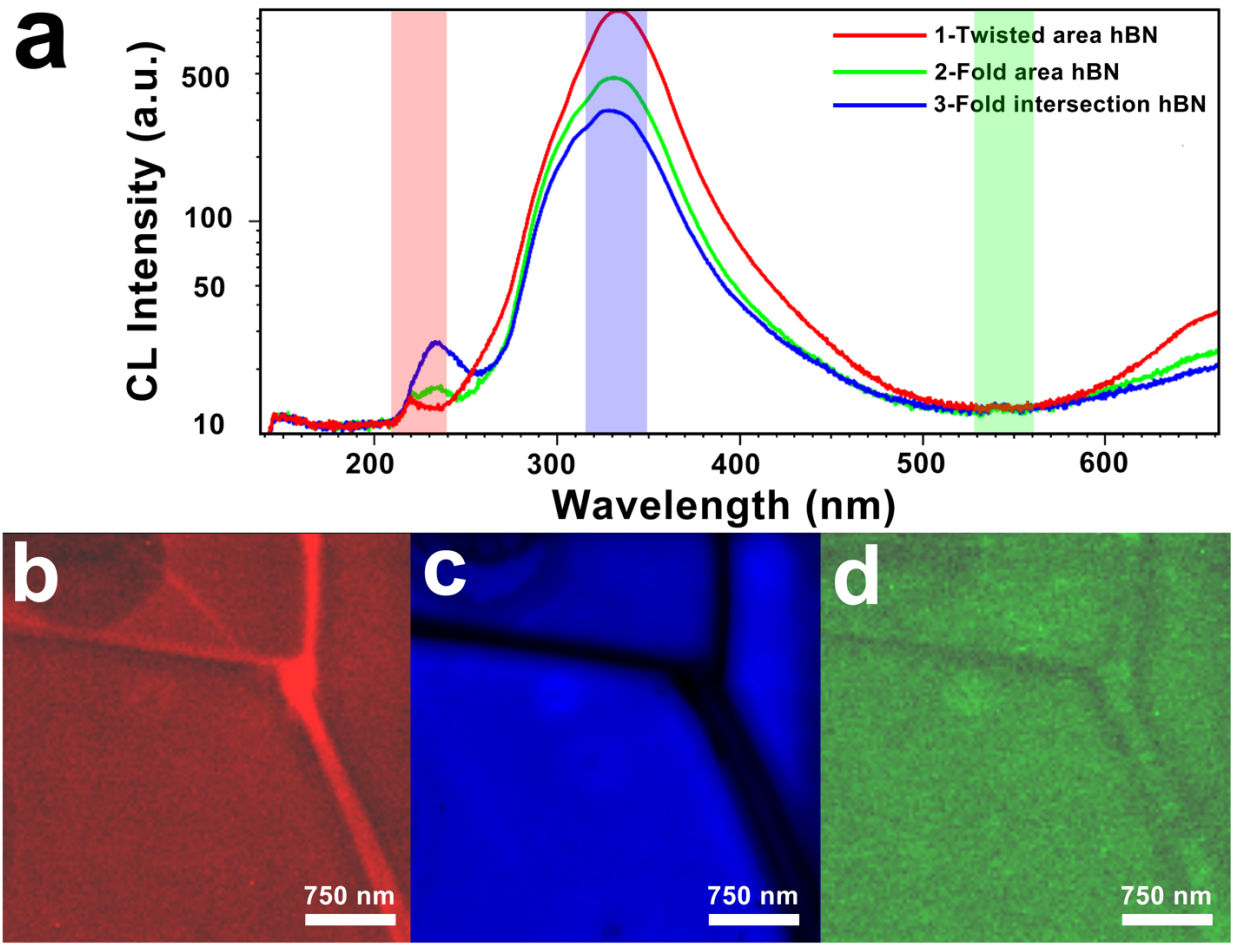

Fig. S3. Colorized bands images of fold area hBN after carbon-coating. (a) CL spectra of hBN. (b) Spectral range is red labeled area in Fig S3(a). (c) Spectral range is blue labeled area in Fig S3(a). (d) Spectral range is green labeled area in Fig S3(a).

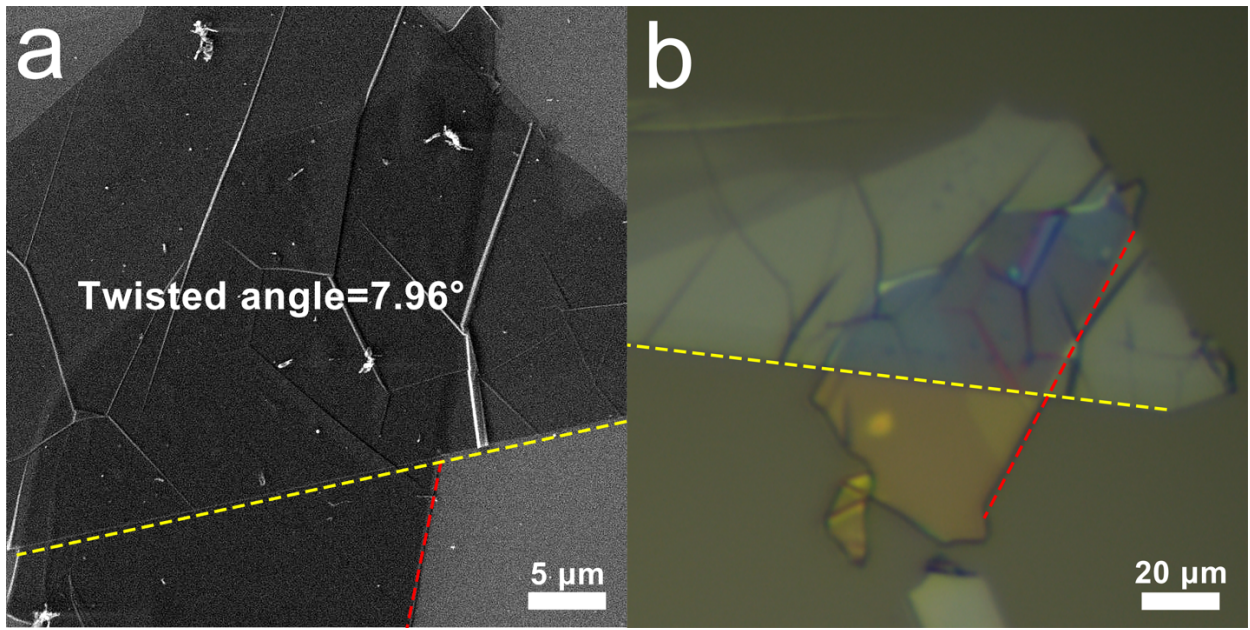

Fig. S4. (a) SE and (b) optical images of twisted hBN. The twisted angle is 7.96°.